\documentclass[a4paper]{article}

\usepackage{amsmath} 
\usepackage{amsthm} 
\usepackage{cite} 
\usepackage{hyperref} 
\usepackage{graphicx}
\usepackage{hyperref}
\usepackage{authblk}
\usepackage[margin=1.2cm]{geometry}
\usepackage{multicol}
\usepackage{blindtext}
\usepackage{epstopdf}
\usepackage{color}
\usepackage[numbib]{tocbibind} 

\usepackage{amssymb}
\usepackage{float}
\usepackage[ruled,linesnumbered]{algorithm2e}
\allowdisplaybreaks[4]
\newtheorem{lemma}{Lemma}{}
\newtheorem{theorem}{Theorem}{}
{}
{}
\newtheorem{assumption}{Assumption}

\begin{document}

\title{Online Low-Carbon Workload, Energy, and Temperature Management of Distributed Data Centers}

\author[1,2]{Rui Xie}
\author[1*]{Yue Chen}
\author[3]{Xi Weng}

\affil[1]{Department of Mechanical and Automation Engineering, The Chinese University of Hong Kong, Hong Kong, China}
\affil[2]{The Shun Hing Institute of Advanced Engineering, The Chinese University of Hong Kong, Hong Kong, China.}
\affil[3]{Guanghua School of Management, Peking University, China. \newline *Email: yuechen@mae.cuhk.edu.hk}
\date{}

\setcounter{Maxaffil}{0}
\renewcommand\Affilfont{\itshape\small}

\maketitle

\begin{changemargin}{1.2cm}{1.2cm} 
\end{changemargin}

\begin{abstract}
Data centers have become one of the major energy consumers, making their low-carbon operations critical to achieving global carbon neutrality. Although distributed data centers have the potential to reduce costs and emissions through cooperation, they are facing challenges due to uncertainties. This paper proposes an online approach to co-optimize the workload, energy, and temperature strategies across distributed data centers, targeting minimal total cost, controlled carbon emissions, and adherence to operational constraints. Lyapunov optimization technique is adopted to derive a parametric real-time strategy that accommodates uncertainties in workload demands, ambient temperature, electricity prices, and carbon intensities, without requiring prior knowledge of their distributions. A theoretical upper bound for the optimality gap is derived, based on which a linear programming problem is proposed to optimize the strategy parameters, enhancing performance while ensuring operational constraints. Case studies and method comparison validate the proposed method's effectiveness in reducing costs and carbon emissions.
\end{abstract}

\begin{multicols}{2}

\section{Introduction}

With the explosive growth of cloud-based services, there has been a surge in the establishment of distributed data centers worldwide. Data centers are crucial for supporting information technology (IT) infrastructure; however, their significant energy consumption also exerts additional environmental pressure. Data center energy consumption is expected to become one of the major components of global electricity demand by 2030 \cite{mytton2022sources}. Therefore, the low-carbon operation of data centers plays an important role in pursuing the goal of carbon neutrality \cite{liu2023optimal}.

Some research works tried to reduce the energy consumption and operation cost of a data center by scheduling workloads. For example, a mixed duration task assignment and migration algorithm was developed in \cite{lou2023energy} based on deep reinforcement learning for energy-efficient purposes in data centers. A data and energy migration optimization method was proposed in \cite{da2023optimization} for the green operation of a mini data center. A hybrid quantum-classical energy management method was introduced in \cite{zhao2024optimal} to minimize the operation cost of a data center.

Recently, distributed data centers have drawn attention, which could be coordinated to further reduce the total operation cost. A day-ahead and intraday scheduling method for Internet data centers was proposed in \cite{ye2023joint}, considering the power losses of uninterruptible power supply. A multi-objective optimization problem was formulated in \cite{khalid2023dual} for the operation of geo-distributed data centers under time-of-use electricity prices, solved by an evolutionary algorithm. The coordination of data centers and their participation in the electricity market were jointly optimized in \cite{sun2020workload} and solved by a polytope cutting algorithm. However, the aforementioned operation strategies either adopt a deterministic paradigm or rely on the predictions for uncertainties, jeopardizing their performances in a strongly uncertain environment without effective predictions.

We accommodate strong uncertainties in the optimization problem by using Lyapunov optimization technique \cite{neely2010stochastic}, which is a prediction-free stochastic programming technique that applies to infinite-time problems with time average objective functions and constraints.
Lyapunov optimization requires no prior knowledge of distributions and can provide theoretical guarantees for the optimality gap of the obtained strategy. Therefore, it shows strength in various scenarios such as energy sharing \cite{liu2018online}, mobile cloud offloading \cite{li2022lyapunov}, and market bidding \cite{xie2024real}.

Lyapunov optimization was applied to the workload and energy management of data centers in \cite{sun2023battery} and then improved in \cite{huang2024online}. However, there are still research gaps: 1) Although the total operation cost is minimized by these methods, there are no direct limits for the carbon emissions, leading to high-emission low-cost strategies in some scenarios. 2) These methods neglect the cooling demand from temperature management, which actually constitutes a non-negligible part of the data centers' energy consumption \cite{mytton2022sources}. 3) The parameter decisions in \cite{sun2023battery} and \cite{huang2024online} rely on careful theoretical deduction, making it difficult to consider emission limits and temperature constraints further.

Aiming to fill the above research gaps, this paper proposes a Lyapunov optimization-based online coordination method for the low-carbon operation of distributed data centers, which jointly manages workload, energy, and temperature. The novelty of this paper is two-fold:

1) The proposed method improves traditional prediction-free distributed data center coordination methods \cite{sun2023battery,huang2024online} by considering emission limits and temperature management. The proposed method is able to control the average emission according to the requirement and maintain the proper temperature using a cooling system, under the uncertainties of workload demands, ambient temperature, electricity prices, and carbon intensities.

2) A linear programming (LP) problem is developed to determine the parameters of the Lyapunov optimization-based parametric operation strategy, which minimizes the optimality gap of the strategy while guaranteeing operational feasibility. Unlike the traditional parameter determination methods that rely on substantial theoretical analysis \cite{sun2023battery,huang2024online}, the proposed approach is easier to use and more suitable for complex problems such as the considered one.

The rest of this paper is organized as follows: The system structure is introduced in Section~\ref{sec:structure}. The problem formulation and the proposed online algorithm are developed in Section~\ref{sec:offline} and Section~\ref{sec:online}, respectively. Case study is presented in Section~\ref{sec:case}. Finally, Section~\ref{sec:conclusion} concludes the paper.

\section{System Structure}
\label{sec:structure}

\begin{figure}[H]
\centering
\includegraphics[width=1\linewidth]{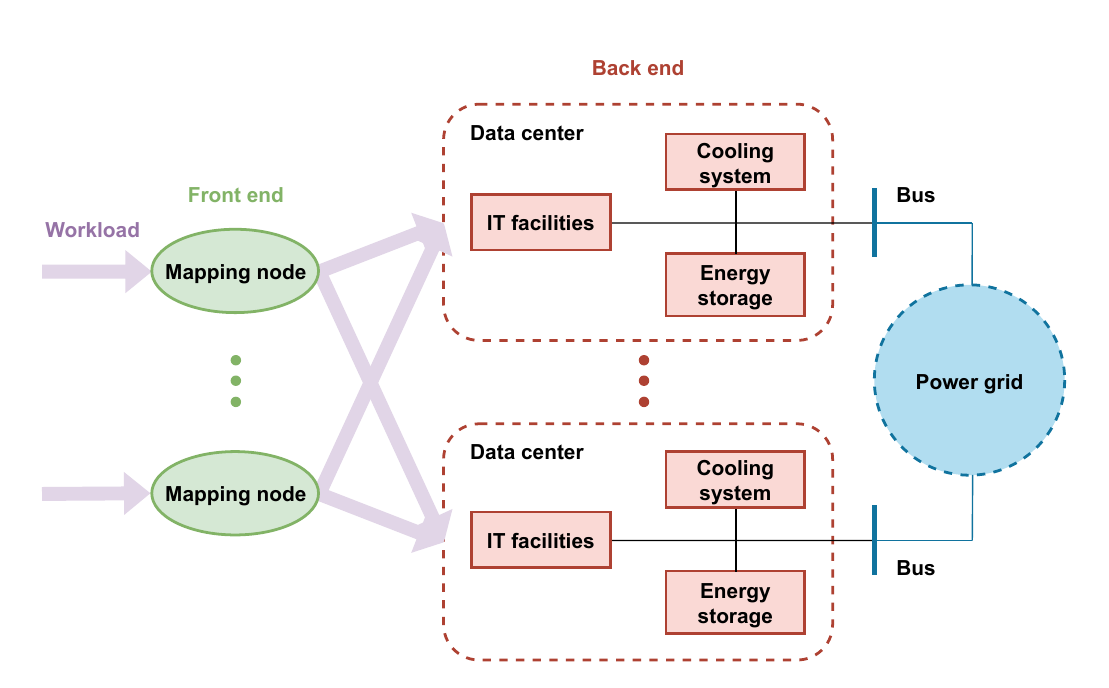}
\caption{The structure of the distributed data center system.}
\label{fig:structure}
\end{figure}

The structure of the distributed data center system is illustrated in Figure~\ref{fig:structure}, which consists of the front end and the back end. The workload demand first arrives at the mapping nodes in the front end. Then they queue here waiting for transfer to the back end. After arriving at data centers in the back end, the workloads queue before they are processed by the IT facilities in the data center. The queues in the front and back ends are both served in the first-in-first-out fashion. Besides IT facilities, there is also an energy storage system and a cooling system in each data center. Data centers are connected to the power grid at different buses.

\section{Problem Formulation}
\label{sec:offline}

In this section, we first establish the component model of the distributed data center system, based on which the offline optimization problem is provided. Finally, the stochastic program that we want to solve is formulated.

\subsection{Component Model}

The workload flow, energy storage system, cooling system, electricity cost, and carbon emission are modeled one by one in this subsection.

Let $\mathcal{T}$, $\mathcal{I}$, and $\mathcal{J}$ be the index sets of the time slots, mapping nodes, and data centers, respectively. Then for any time slot $t \in \mathcal{T}$, the workload flow constraints are as follows:
\begin{subequations}
\label{eq:workload}
\begin{align}
    \label{eq:workload-c}
    & 0 \leq a_{i t}^F \leq \alpha_{i t}^F, \forall i \in \mathcal{I}, \\
    \label{eq:workload-d}
    & 0 \leq m_{i j t}^R \leq M_{i j}^R, \forall i \in \mathcal{I}, \forall j \in \mathcal{J}, \\
    \label{eq:workload-e}
    & 0 \leq p_{j t}^B \leq P_j^B, \forall j \in \mathcal{J}, \\
    \label{eq:workload-f}
    & q_{i t}^F \geq 0, \forall i \in \mathcal{I}, q_{j t}^B \geq 0, \forall j \in \mathcal{J}, \\
    \label{eq:workload-a}
    & q_{i (t+1)}^F = q_{i t}^F + a_{i t}^F - \sum_{j \in \mathcal{J}} m_{i j t}^R, \forall i \in \mathcal{I}, \\
    \label{eq:workload-b}
    & q_{j (t+1)}^B = q_{j t}^B + \sum_{i \in \mathcal{I}} m_{i j t}^R - p_{j t}^B, \forall j \in \mathcal{J}.
\end{align}
\end{subequations}
In \eqref{eq:workload-c}, $\alpha_{i t}^F$ denotes the workload demand arriving at mapping node $i$ in time slot $t$, which is uncertain; $a_{i t}^F$ is the workload demand that mapping node $i$ accepts. In \eqref{eq:workload-d}, $m_{i j t}^R$ is the workload transferred from mapping node $i$ to data center $j$, whereas $M_{i j}^R$ denotes the capacity of the link from $i$ to $j$. In \eqref{eq:workload-e}, $p_{j t}^B$ is the workload processed at data center $j$ in time slot $t$, which should be no larger than the capacity of the IT facilities at $j$, denoted by $P_j^B$. Constraint \eqref{eq:workload-f} stipulates the nonnegativity of the queues in the front and back ends, denoted by $q_{i t}^F$ and $q_{j t}^B$, respectively. Constraints \eqref{eq:workload-a} and \eqref{eq:workload-b} describe the dynamics of the queues. The operation cost caused by the workload flow is modeled as follows:
\begin{align}
    \label{eq:workload-cost}
    f_t^W = \sum_{i \in \mathcal{I}} \sum_{j \in \mathcal{J}} \gamma_{i j}^R m_{i j t}^R + \sum_{i \in \mathcal{I}} \gamma_i^F (\alpha_{i t}^F - a_{i t}^F),
\end{align}
where the first term is the workload transfer cost, the second term is the penalty for rejecting workload demand, and $\gamma_{i j}^R$ and $\gamma_i^F$ are cost coefficients.

The constraints of the energy storage system in time slot $t$ are as follows:
\begin{subequations}
\label{eq:ES}
\begin{align}
    \label{eq:ES-b}
    & 0 \leq p_{j t}^{SC} \leq P_j^{SC}, 0 \leq p_{j t}^{SD} \leq P_j^{SD}, \forall j \in \mathcal{J}, \\
    \label{eq:ES-d}
    & \underline{E}_j^S \leq e_{j t}^S \leq \overline{E}_j^S, \forall j \in \mathcal{J}, \\
    \label{eq:ES-a}
    & e_{j (t + 1)}^S = e_{j t}^S + p_{j t}^{SC} \eta_j^{SC} - p_{j t}^{SD} / \eta_j^{SD}, \forall j \in \mathcal{J}.
\end{align}
\end{subequations}
In \eqref{eq:ES}, $p_{j t}^{SC}$ and $p_{j t}^{SD}$ are the charging and discharging power, respectively. The stored energy at the beginning of time slot $t$ is denoted by $e_{j t}^S$. Therefore, constraints \eqref{eq:ES-b} and \eqref{eq:ES-d} stipulate bounds for these variables. Constraint \eqref{eq:ES-a} models the change of stored energy, where $\eta_j^{SC}$ and $\eta_j^{SD}$ are the efficiency coefficients. To consider battery degradation, the operation cost of the energy storage systems is formulated as follows:
\begin{align}
    \label{eq:ES-cost}
    f_t^S = \sum_{j \in \mathcal{J}} \gamma_j^S (p_{j t}^{SC} + p_{j t}^{SD}).
\end{align}

The cooling system and the temperature control of the data center are modeled as follows:
\begin{subequations}
\label{eq:cooling}
\begin{align}
    \label{eq:cooling-a}
    & 0 \leq p_{j t}^C \leq P_j^C, \forall j \in \mathcal{J}, \\
    \label{eq:cooling-b}
    & \underline{T}_j^H \leq \tau_{j t}^H \leq \overline{T}_j^H, \forall j \in \mathcal{J}, \\
    \label{eq:cooling-c}
    & \tau_{j (t + 1)}^H = \tau_{j t}^H + \kappa_j^B p_{j t}^B - \kappa_j^C p_{j t}^C - \beta_{j t}^C, \forall j \in \mathcal{J}.
\end{align}
\end{subequations}
Constraint \eqref{eq:cooling-a} bounds the power of the cooling system. Constraint \eqref{eq:cooling-b} is for the limits of the temperature $\tau_{j t}^H$ in the data center. In \eqref{eq:cooling-c}, the temperature changes are modeled, where the IT facilities increase the temperature and the cooling system lowers it with coefficients $\kappa_j^B$ and $\kappa_j^C$, respectively. The ambient temperature also affects $\tau_{j t}^H$, whose effect is represented by the uncertain term $\beta_{j t}^C$. The operation cost of cooling system is primarily driven by electricity usage, which is included in \eqref{eq:electricity-cost}.

The power demand of data center $j$ is:
\begin{align}
    \label{eq:electricity}
    p_{j t}^G = p_{j t}^B + p_{j t}^{SC} - p_{j t}^{SD} + p_{j t}^C, \forall j \in \mathcal{J}.
\end{align}
Therefore, the electricity cost is:
\begin{align}
    \label{eq:electricity-cost}
    f_t^G = \sum_{j \in \mathcal{J}} \gamma_{j t}^G p_{j t}^G,
\end{align}
where the electricity price $\gamma_{j t}^G$ is uncertain. At the same time, the power demand causes carbon emissions in the power generation process. Due to the randomness of renewable energy generation, the carbon intensity $\gamma_{j t}^E$ is also uncertain across time. To ensure low-carbon operation, we use the following constraint to upper bound the average carbon emission rate:
\begin{align}
    \label{eq:emission}
    \frac{1}{T} \sum_{t \in \mathcal{T}} \sum_{j \in \mathcal{J}} \gamma_{j t}^E p_{j t}^G \leq C^E,
\end{align}
where $T$ is the number of time slots.

\subsection{Offline Optimization Problem}

Combining the models of all the components and minimizing the total operation cost, we have the following offline optimization problem for the distributed data center coordination:
\begin{subequations}
\label{eq:offline}
\begin{align}
    \min~ & \frac{1}{T} \sum_{t \in \mathcal{T}} (f_t^W + f_t^S + f_t^G), \\
    \text{s.t.}~ & \eqref{eq:workload}\text{--}\eqref{eq:electricity-cost}, \forall t \in \mathcal{T}, \eqref{eq:emission}.
\end{align}
\end{subequations}
Problem \eqref{eq:offline} is an LP problem. However, there are uncertainties $\alpha_{i t}^F$, $\beta_{j t}^C$, $\gamma_{j t}^G$, and $\gamma_{j t}^E$ that we cannot know in advance, making the direct solution of problem \eqref{eq:offline} impractical. 

\subsection{Stochastic Formulation}

At time slot $t$, we know the uncertainty realizations at and before time slot $t$. This paper aims to propose a prediction-free strategy, which means that we decide the strategy at each time slot according to the current and previous uncertainty realizations as well as system status. In the following, we only consider such prediction-free strategies.

To consider the long-term expected effect, we change the time horizon to infinite, i.e., $\mathcal{T} = \mathbb{N}_+ = \{1, 2, \dots, t \dots \}$. Then the accumulated values over time in the objective function and constraint \eqref{eq:emission} should be replaced by time average terms. In addition, the workload queues should be controlled to avoid linear increase. Therefore, the problem is changed into:
\begin{subequations}
\label{eq:stochastic}
\begin{align}
    & \min~ \lim_{T \rightarrow \infty} \frac{1}{T} \sum_{t = 1}^T \mathbb{E} [f_t^W + f_t^S + f_t^G], \\
    & \text{s.t.}~ \eqref{eq:workload}\text{--}\eqref{eq:electricity-cost}, \forall t, \\
    \label{eq:stochastic-c}
    & \lim_{T \rightarrow \infty} \frac{\mathbb{E}[\tilde{q}_{i T}^F]}{T} = 0, \forall i \in \mathcal{I}, \lim_{T \rightarrow \infty} \frac{\mathbb{E}[\tilde{q}_{j T}^B]}{T} = 0, \forall j \in \mathcal{J}, \\
    \label{eq:stochastic-d}
    & \lim_{T \rightarrow \infty} \frac{1}{T} \sum_{t = 1}^T \sum_{j \in \mathcal{J}} \mathbb{E}[\gamma_{j t}^E p_{j t}^G] \leq C^E,
\end{align}
\end{subequations}
where $\mathbb{E}[\cdot]$ means taking expectation. Problem \eqref{eq:stochastic} is difficult to solve due to the uncertainties and the infinite time horizon. We need an online optimization algorithm to guarantee feasibility and ensure good optimality in the infinite-time case.

\section{Online Optimization Algorithm}
\label{sec:online}

This section introduces the proposed online optimization algorithm. A parametric online algorithm is developed to provide the operation strategy using the Lyapunov optimization technique. Then an LP-based method is proposed to optimize the parameters of the strategy. Meanwhile, feasibility guarantee and optimality gaps are analyzed and proven.

\subsection{Parametric Online Algorithm}

Following the theory of Lyapunov optimization \cite{neely2010stochastic}, we first define virtual queues $\tilde{q}_{i t}^F$, $\tilde{q}_{j t}^B$, $\tilde{q}_{j t}^S$, $\tilde{q}_{j t}^H$, and $\tilde{q}_t^E$: For any $t \in \mathbb{N}_+$,
\begin{subequations}
\label{eq:virtual-queue}
\begin{align}
    & \tilde{q}_{i t}^F = q_{i t}^F + \theta_i^F, \forall i \in \mathcal{I}, \\
    & \tilde{q}_{j t}^B = q_{j t}^B + \theta_j^B, \tilde{q}_{j t}^S = e_{j t}^S + \theta_j^S, \tilde{q}_{j t}^H = \tau_{j t}^H + \theta_j^H, \forall j \in \mathcal{J}, \\
    \label{eq:virtual-queue-E}
    & \tilde{q}_{t + 1}^E = \max \left\{ \tilde{q}_t^E + \sum_{j \in \mathcal{J}} \gamma_{j t}^E p_{j t}^G - C^E, 0 \right\},
\end{align}
\end{subequations}
where $\theta_i^F$, $\theta_j^B$, $\theta_j^S$, and $\theta_j^H$ are parameters to be decided later; the initial state of $\tilde{q}^E$ is $\tilde{q}_1^E = 0$. In Lyapunov optimization, we require the virtual queues to be mean rate stable, i.e.,
\begin{subequations}
\label{eq:mean-rate-stable}
\begin{align}
    \label{eq:mean-rate-stable-b}
    & \eqref{eq:stochastic-c}, \lim_{T \rightarrow \infty} \frac{\mathbb{E}[\tilde{q}_{j T}^S]}{T} = 0, \lim_{T \rightarrow \infty} \frac{\mathbb{E}[\tilde{q}_{j T}^H]}{T} = 0, \forall j \in \mathcal{J}, \\
    \label{eq:mean-rate-stable-c}
    & \lim_{T \rightarrow \infty} \mathbb{E}[\tilde{q}_T^E] / T = 0.
\end{align}
\end{subequations}

Lemma~\ref{lemma:mean-rate-stable} is about the relationships between \eqref{eq:mean-rate-stable} and the constraints in \eqref{eq:stochastic}, whose proof can be found in the appendix.
\begin{lemma}
\label{lemma:mean-rate-stable}
    Feasible solutions of problem \eqref{eq:stochastic} satisfy \eqref{eq:mean-rate-stable-b}. Constraint \eqref{eq:mean-rate-stable-c} implies \eqref{eq:stochastic-d}.
\end{lemma}

To apply the Lyapunov optimization technique, we relax the bound constraints \eqref{eq:workload-f}, \eqref{eq:ES-d}, and \eqref{eq:cooling-b} of queues, revise \eqref{eq:stochastic-d}, and replace them by constraint \eqref{eq:mean-rate-stable}. Then we have the following mean rate optimization problem:
\begin{subequations}
\label{eq:mean}
\begin{align}
    & F^m = \min~ \lim_{T \rightarrow \infty} \frac{1}{T} \sum_{t = 1}^T \mathbb{E} [f_t^W + f_t^S + f_t^G] \\
    \label{eq:mean-a}
    & \text{s.t.}~ \tilde{q}_{i (t+1)}^F = \tilde{q}_{i t}^F + a_{i t}^F - \sum_{j \in \mathcal{J}} m_{i j t}^R, \forall i, \forall t, \\
    & \tilde{q}_{j (t+1)}^B = \tilde{q}_{j t}^B + \sum_{i \in \mathcal{I}} m_{i j t}^R - p_{j t}^B, \forall j, \forall t, \\
    & \tilde{q}_{j (t + 1)}^S = \tilde{q}_{j t}^S + p_{j t}^{SC} \eta_j^{SC} - p_{j t}^{SD} / \eta_j^{SD}, \forall j, \forall t, \\
    & \tilde{q}_{j (t + 1)}^H = \tilde{q}_{j t}^H + \kappa_j^B p_{j t}^B - \kappa_j^C p_{j t}^C - \beta_{j t}^C, \forall j, \forall t, \\
    \label{eq:mean-f}
    & \tilde{q}_{t + 1}^E = \max \left\{ \tilde{q}_t^E + \sum_{j \in \mathcal{J}} \gamma_{j t}^E p_{j t}^G - C^E, 0 \right\}, \forall t, \\
    & \eqref{eq:workload-c}\text{--}\eqref{eq:workload-e}, \eqref{eq:workload-cost}, \eqref{eq:ES-b}, \eqref{eq:ES-cost}, \eqref{eq:cooling-a}, \eqref{eq:electricity}, \eqref{eq:electricity-cost}, \forall t, \eqref{eq:mean-rate-stable}.
\end{align}
\end{subequations}

Following the Lyapunov optimization theory, the Lyapunov function is as follows:
\begin{align}
    L_t = \frac{\sum_{i \in \mathcal{I}} (\tilde{q}_{i t}^F)^2 + \sum_{j \in \mathcal{J}} ((\tilde{q}_{j t}^B)^2 + (\tilde{q}_{j t}^S)^2 + (\tilde{q}_{j t}^H)^2) + (\tilde{q}_t^E)^2}{2}, \forall t. \nonumber
\end{align}
The Lyapunov drift is defined as $\Delta_t = L_{t + 1} - L_t$. To derive an upper bound for $\Delta_t$, we need assumptions about the range of uncertainties:
\begin{assumption}
\label{assumption:bound}
    $0 \leq \alpha_{i t}^F \leq A_i^F$, $\forall i \in \mathcal{I}$, $\forall t$. The data center $j$'s ambient temperature is no lower than $\underline{T}_j^H$, $\forall j \in \mathcal{J}$. $\underline{\beta}_j^C \leq \beta_{j t}^C \leq \overline{\beta}_j^C$, $\underline{\gamma}_j^G \leq \gamma_{j t}^G \leq \overline{\gamma}_j^G$, $0 \leq \underline{\gamma}_j^E \leq \gamma_{j t}^E \leq \overline{\gamma}_j^E$, $\forall j$, $\forall t$.
\end{assumption}

Lemma~\ref{lemma:drift} gives an upper bound for $\Delta_t$, which is a linear function of the decision variables in time slot $t$ (note that the queues $\tilde{q}_{i t}^F$, $\tilde{q}_{j t}^B$, $\tilde{q}_{j t}^S$, $\tilde{q}_{j t}^H$, and $\tilde{q}_t^E$ have been decided before time slot $t$). The proof of Lemma~\ref{lemma:drift} is in the appendix.
\begin{lemma}
\label{lemma:drift}
    Suppose that Assumption~\ref{assumption:bound} holds. Then $\Delta_t \leq I + B$, where $B$ is a constant independent of $t$ and
    \begin{align}
    \label{eq:I}
    & I = \sum_{i \in \mathcal{I}} \tilde{q}_{i t}^F \left( a_{i t}^F - \sum_{j \in \mathcal{J}} m_{i j t}^R \right) + \sum_{j \in \mathcal{J}} \tilde{q}_{j t}^B \left( \sum_{i \in \mathcal{I}} m_{i j t}^R - p_{j t}^B \right) \nonumber \\
    & + \sum_{j \in \mathcal{J}} \tilde{q}_{j t}^S ( p_{j t}^{SC} \eta_j^{SC} - p_{j t}^{SD} / \eta_j^{SD} ) + \sum_{j \in \mathcal{J}} \tilde{q}_{j t}^H ( \kappa_j^B p_{j t}^B - \kappa_j^C p_{j t}^C - \beta_{j t}^C ) \nonumber \\
    & + \tilde{q}_t^E \left( \sum_{j \in \mathcal{J}} \gamma_{j t}^E (p_{j t}^B + p_{j t}^{SC} - p_{j t}^{SD} + p_{j t}^C) - C^E \right).
\end{align}
\end{lemma}

The online optimization problem in time slot $t$ minimizes the weighted sum of the Lyapunov drift's upper bound and the operation cost in the current time slot, which is as follows:
\begin{subequations}
\label{eq:online}
\begin{align}
    \min~ & I + V (f_t^W + f_t^S + f_t^G) \\
    \text{s.t.}~ & \eqref{eq:workload-c}\text{--}\eqref{eq:workload-e}, \eqref{eq:workload-cost}, \eqref{eq:ES-b}, \eqref{eq:ES-cost}, \eqref{eq:cooling-a}, \eqref{eq:electricity}, \eqref{eq:electricity-cost},
\end{align}
\end{subequations}
where $V \geq 0$ is a parameter to be determined. Given parameters $V$, $\theta_i^F$, $\theta_j^B$, $\theta_j^S$, and $\theta_j^H$, the steps of the proposed online algorithm are presented in Algorithm~\ref{alg:online}.
\begin{algorithm}[H]
\label{alg:online}
\DontPrintSemicolon 
\SetAlgoLined
Observe the uncertainty realizations of $\alpha_{i t}^F$, $\beta_{j t}^C$, $\gamma_{j t}^G$, and $\gamma_{j t}^E$.\;
Solve the LP problem \eqref{eq:online} and obtain the values of $a_{i t}^F$, $m_{i j t}^R$, $p_{j t}^B$, $p_{j t}^{SC}$, $p_{j t}^{SD}$, and $p_{j t}^C$.\;
Update the queues according to \eqref{eq:mean-a}--\eqref{eq:mean-f} and obtain $\tilde{q}_{i (t + 1)}^F$, $\tilde{q}_{j (t + 1)}^B$, $\tilde{q}_{j (t + 1)}^S$, $\tilde{q}_{j (t + 1)}^H$, and $\tilde{q}_{t + 1}^E$.\;
\caption{Online algorithm in time slot $t$.}
\end{algorithm}

Algorithm~\ref{alg:online} is parametric. We denote its performance by $F^l(V, \theta)$, which equals $\lim_{T \rightarrow \infty} \frac{1}{T} \sum_{t = 1}^T \mathbb{E} [f_t^W + f_t^S + f_t^G]$ and depends on the choices of $V$ and $\theta = (\theta_i^F, \theta_j^B, \theta_j^S, \theta_j^H)$.

\subsection{Performance Guarantee and Parameter Optimization}

To analyze the performance guarantee, we assume that the uncertainties are independent and identically distributed:
\begin{assumption}
\label{assumption:iid}
    $\xi_t = (\alpha_{i t}^F, i \in \mathcal{I}, \beta_{j t}^C, \gamma_{j t}^G, \gamma_{j t}^E, j \in \mathcal{J}), \forall t \in \mathcal{T}$ are independent and identically distributed.
\end{assumption}

Theorem~\ref{thm:performance} gives a performance guarantee for the parametric online algorithm, which follows from Theorem 4.8 in \cite{neely2010stochastic}.
\begin{theorem} 
\label{thm:performance}
    Suppose that Assumption~\ref{assumption:bound} and Assumption~\ref{assumption:iid} hold. With parameter $V > 0$ and arbitrary $\theta$, the optimality gap of the online algorithm is upper bounded by:
    \begin{align}
        F^l(V, \theta) - F^m \leq B / V. \nonumber
    \end{align}
\end{theorem}

According to Theorem~\ref{thm:performance}, the larger $V$, the better the performance. However, since relaxations are adopted while deriving problem \eqref{eq:mean}, we should properly choose the parameters to ensure the feasibility of the strategy in problem \eqref{eq:stochastic}. To analyze the feasibility, we consider the following assumption:

\begin{assumption}
\label{assumption:QE}
    $\tilde{q}_t^E \leq Q^E$, $\forall t$. $\kappa_j^C P_j^C \geq \kappa_j^B P_j^B - \underline{\beta}_j^C, \forall j \in \mathcal{J}$ always holds, i.e., the cooling systems' capacities are large enough to prevent the temperature increase in all cases.
\end{assumption}

Theorem~\ref{thm:feasibility} provides the constraints that can guarantee the feasibility, whose proof is in the appendix.
\begin{theorem}
\label{thm:feasibility}
    Suppose that Assumption~\ref{assumption:bound} and Assumption~\ref{assumption:QE} hold and the parameters $V \geq 0$ and $\theta$ satisfy \eqref{eq:requirement}. Then the strategy by Algorithm~\ref{alg:online} is feasible in problem \eqref{eq:stochastic}:
    \begin{subequations}
    \label{eq:requirement}
    \begin{align}
        & - \theta_i^F - \sum_{j' \in \mathcal{J}} M_{i j'}^R + \theta_j^B + V \gamma_{i j}^R \geq 0, \forall i \in \mathcal{I}, \forall j \in \mathcal{J}, \\
        & - \theta_j^B - P_j^B + (\underline{T}_j^H + \theta_j^H) \kappa_j^B + V \underline{\gamma}_j^G \geq 0, \forall j \in \mathcal{J}, \\
        & - \frac{\theta_j^S + \underline{E}_j^S + P_j^{SD}/\eta_j^{SD}}{\eta_j^{SD}} - Q^E \overline{\gamma}_j^E + V \gamma_j^S - V \overline{\gamma}_j^G \geq 0, \forall j \in \mathcal{J}, \\
        & (\theta_j^S + \overline{E}_j^S - P_j^{SC} \eta_j^{SC}) \eta_j^{SC} + V \gamma_j^S + V \underline{\gamma}_j^G \geq 0, \forall j \in \mathcal{J}, \\
        & - (\theta_j^H + \underline{T}_j^S + \kappa_j^C P_j^C) \kappa_j^C + V \underline{\gamma}_j^G \geq 0, \forall j \in \mathcal{J}, \\
        & - (\theta_j^H + \overline{T}_j^S - \kappa_j^B P_j^B + \underline{\beta}_j^C) \kappa_j^C \nonumber \\
        & + Q^E \overline{\gamma}_j^E + V \overline{\gamma}_j^G \leq 0, \forall j \in \mathcal{J}.
    \end{align}
    \end{subequations}
\end{theorem}

According to Theorem~\ref{thm:performance} and Theorem~\ref{thm:feasibility}, the following LP problem is proposed to optimize $V$ and $\theta$:
\begin{align}
\label{eq:optimize-parameter}
    \max_{V, \theta}~ & V ~\text{s.t.}~ \eqref{eq:requirement}.
\end{align}
Although the bounds of uncertainties can be easily estimated by the historical data, the value $Q^E$ is difficult to know in advance. Therefore, Algorithm~\ref{alg:parameter} is developed to determine $Q^E$ and $(V, \theta)$ in an iterative manner.
\begin{algorithm}[H]
\label{alg:parameter}
\DontPrintSemicolon 
\SetAlgoLined
\SetKwProg{Process}{}{:}{}{}
Let $Q^E = 0$.\;
\Process{Repeat}{
Solve the LP problem \eqref{eq:optimize-parameter} and obtain $(V, \theta)$.\;
Simulate the online algorithm using the historical dataset and let $Q^E = \max_t \tilde{q}_t^E$.\;
}{Until $Q^E$ converges.}
\caption{Parameter optimization algorithm.}
\end{algorithm}

\section{Case Study}
\label{sec:case}

In this section, we test the proposed method using a numerical simulation. The settings and results of the proposed method are shown in Section~\ref{sec:case-base}, method comparison is reported in Section~\ref{sec:case-comparison}, and parameter sensitivity analysis is conducted in Section~\ref{sec:case-psa}. All the algorithms are implemented on a laptop with Intel(R) Core(TM) i5-1335U and 16 GB RAM. The LP problems are solved by Gurobi 11.0.3. The code and data can be found in \cite{xie2024github}.

\subsection{Settings and Results}
\label{sec:case-base}

\begin{figure}[H]
\centering
\includegraphics[width=0.7\linewidth]{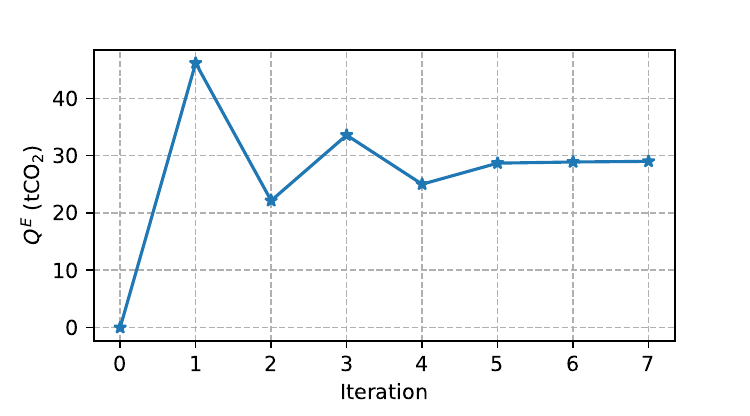}
\caption{The value of $Q^E$ in the iteration process.}
\label{fig:iteration}
\end{figure}

In the test case, there are 2 mapping nodes and 3 data centers. The upper bound $C^E$ of emission rate is set as 1.2 tCO$_2$/h. The historical data of uncertainties are generated from probability distributions, which include 10,000 time slots in total. We first use the 1,000 time slots of data to determine the parameters $(V, \theta)$. The iteration process of the proposed method is drawn in Figure~\ref{fig:iteration}, which shows that Algorithm~\ref{alg:parameter} successfully converges after about 5 iterations.

\begin{figure}[H]
\centering
\includegraphics[width=0.9\linewidth]{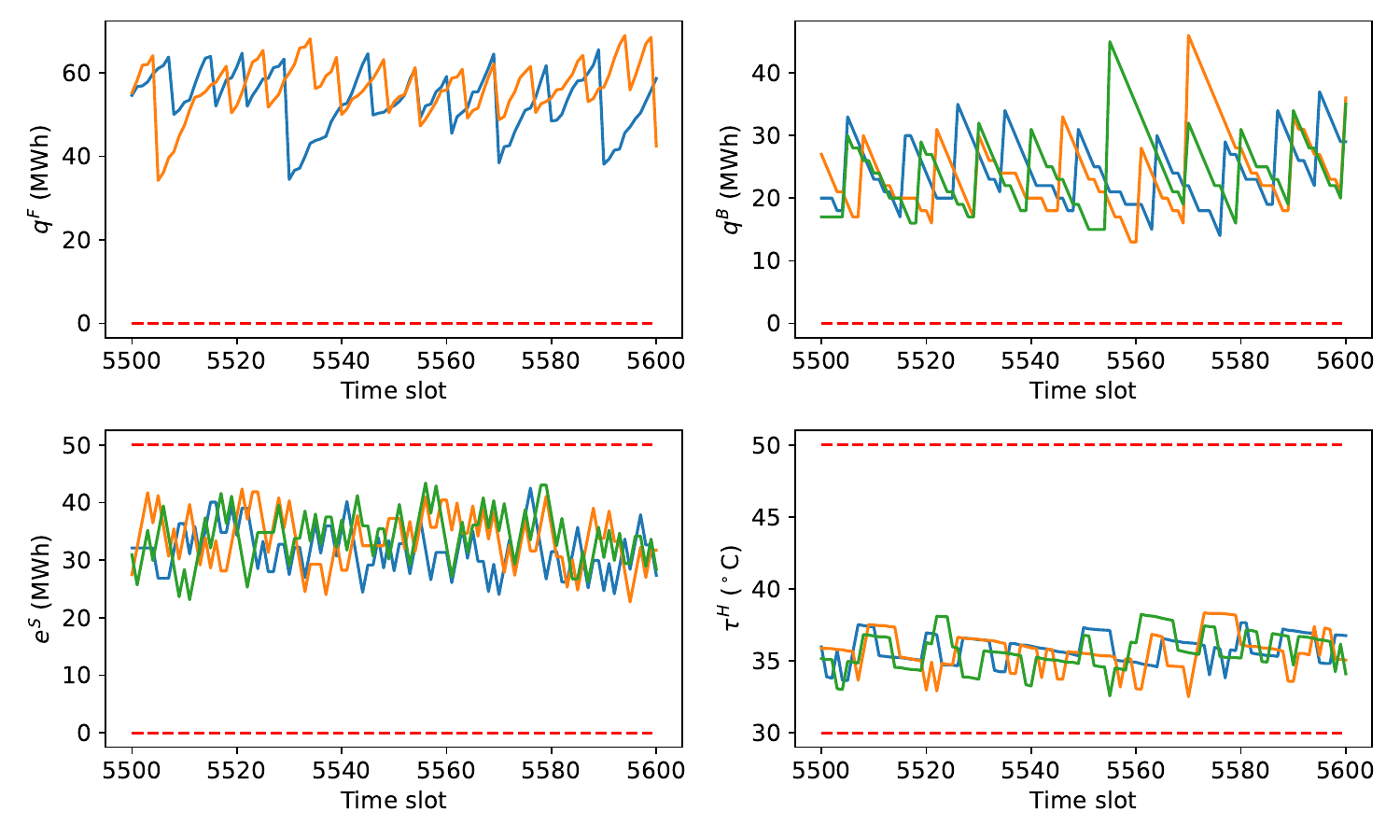}
\caption{The simulated queues $q^F$, $q^B$, $e^S$, and $\tau^H$ in time slots 5,500--5,600. The bounds are represented by red dashed lines. All the queues are within the bounds.}
\label{fig:series}
\end{figure}

The remaining 9,000 time slots are used for performance and feasibility tests. The test average cost rate is 335.3 \$/h. The emission rate is 1.160 tCO$_2$/h, which is slightly lower than the specified upper bound ($C^E$ = 1.2 tCO$_2$/h). Thus, the proposed method is effective in limiting the emissions. In addition, all the simulated queues are within their bounds, which verifies that the proposed method ensures feasibility. Part of the simulated queues are depicted in Figure~\ref{fig:series}.

\subsection{Method Comparison}
\label{sec:case-comparison}

To further verify the effectiveness of the proposed method, we compare it with other methods:

C1 (offline \& low-carbon): Assume the knowledge of future uncertainty realizations and solve the offline problem \eqref{eq:offline}.

C2 (greedy \& low-carbon): Minimize the total operation cost in the current time slot.

C3 (offline): The offline method with no emission bound.

C4 (greedy): The greedy method with no emission bound.

C5 (no emission bound): Use the proposed method but do not consider the emission bound.

In the offline and greedy methods C1--C4, upper bounds for the front- and back-end queues ($Q^F$ = 90 MWh and $Q^B$ = 70 MWh) are added to avoid endless increase of queues. Although offline methods are impractical, they serve as references and provide lower bounds for the cost of online algorithms. All the methods above provide feasible strategies, but their performances in cost and emissions differ, as compared in Table~\ref{tab:comparison}. The proposed method achieves 133.5\% cost and 96.67\% emissions compared to the offline method C1. While the offline method is impractical, the proposed method is an online algorithm that achieves good performances, much better than the greedy algorithm C2. When the emission bound is not considered, the emission rates in methods C3--C5 become much higher than $C^E$ = 1.2 tCO$_2$/h, which emphasizes the need to consider the emission bound in the online optimization for the low-carbon operation of distributed data centers.
\begin{table}[H]
\renewcommand{\arraystretch}{1.2}
\caption{Method Comparison}
\label{tab:comparison}
\centering
\small
\vspace{0.5em}
\begin{tabular}{c c c}
\hline
Method & Cost rate (\$/h) & Emission rate (tCO$_2$/h) \\ 
\hline
Proposed & 335.3 & 1.160 \\
C1 & 251.2 & 1.200 \\
C2 & 1378 & 1.200 \\
C3 & 231.7 & 2.784 \\
C4 & 457.2 & 2.384\\
C5 & 251.6 & 2.803 \\
\hline
\end{tabular}
\end{table}

\subsection{Parameter Sensitivity Analysis}
\label{sec:case-psa}

\begin{figure}[H]
\centering
\includegraphics[width=0.9\linewidth]{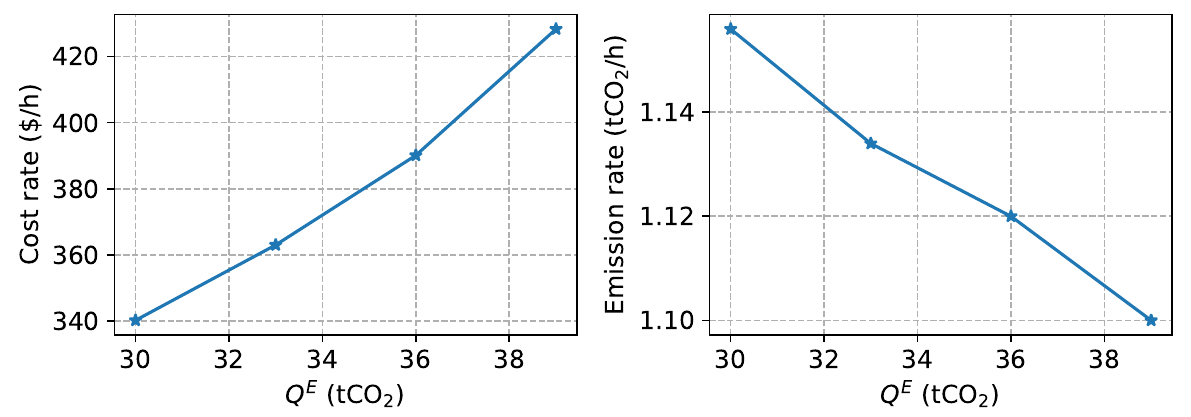}
\caption{Test results under different $Q^E$.}
\label{fig:Q_E}
\end{figure}

We analyze the impacts of the upper bounds $Q^E$ and $C^E$. $Q^E$ is the upper bound of the virtual queue $\tilde{q}^E$, used in the parameter optimization \eqref{eq:optimize-parameter}. The test results under different $Q^E$ values are depicted in Figure~\ref{fig:Q_E}. When $Q^E$ becomes larger, the cost rate increases and the emission rate decreases, where the emission rate is always lower than $C^E$ = 1.2 tCO$_2$/h. Thus, a smaller $Q^E$ is preferred to decrease the cost. This is why we propose Algorithm~\ref{alg:parameter} to find the best $Q^E$.

\begin{figure}[H]
\centering
\includegraphics[width=0.9\linewidth]{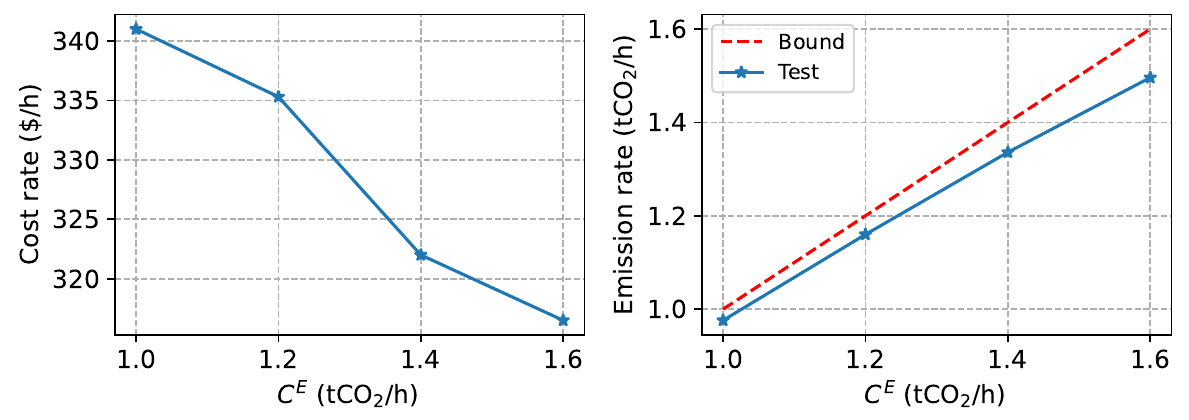}
\caption{Test results under different $C^E$.}
\label{fig:C_E}
\end{figure}

The proposed method is also tested under different $C^E$ values, as drawn in Figure~\ref{fig:C_E}. It shows that the proposed method has lower emission rates than the specified upper bound in all cases, while a lower $C^E$ value leads to a higher cost rate. The results verify the effectiveness of the proposed method in the low-carbon online operation of data centers.



\section{Conclusion}
\label{sec:conclusion}

This paper proposes a novel online low-carbon management method for distributed data centers. While minimizing the total operation cost, the proposed method jointly considers workload, energy, and temperature operational constraints and effectively deals with the uncertainties of workload demands, ambient temperature, electricity prices, and carbon intensities. The proposed method is prediction-free thanks to the adoption of the Lyapunov optimization technique. There are two contributions in this paper: 1) The emission rate upper bound and temperature management are considered in the proposed Lyapunov optimization-based online algorithm. 2) An LP-based method is proposed to optimize the parameters of the operation strategy and to enhance performance while ensuring feasibility. The case study and the method comparison demonstrate the proposed method's effectiveness in low-carbon operation. Incorporating more detailed data center operation models will be our future research direction.

\section{Acknowledgments}

This work was supported by the National Natural Science Foundation of China (Grant No. 72225001), and the Shun Hing Institute of Advanced Engineering, The Chinese University of Hong Kong, through Project RNE-p2-23 (Grant No. 8115071).


\section*{Appendix}

\subsection*{Proof of Lemma~\ref{lemma:mean-rate-stable}}

We first consider a feasible solution to problem \eqref{eq:stochastic}. By constraints \eqref{eq:ES-d} and \eqref{eq:cooling-b}, $\tilde{q}_{j t}^S, \forall t$ and $\tilde{q}_{j t}^H, \forall t$ are bounded, so $\lim_{T \rightarrow \infty} \mathbb{E}[\tilde{q}_{j T}^S] / T = 0$ and $\lim_{T \rightarrow \infty} \mathbb{E}[\tilde{q}_{j T}^H] / T = 0$. Therefore, constraint \eqref{eq:mean-rate-stable-b} is satisfied.

Next, assume that constraint \eqref{eq:mean-rate-stable-c} holds. Because $\tilde{q}_{t + 1}^E \geq \tilde{q}_t^E + \sum_{j \in \mathcal{J}} \gamma_{j t}^E p_{j t}^G - C^E$ and $\tilde{q}_1^E = 0$, we have $\tilde{q}_T^E \geq \sum_{t = 1}^{T - 1} (\sum_{j \in \mathcal{J}} \gamma_{j t}^E p_{j t}^G - C^E)$. Therefore, constraint \eqref{eq:mean-rate-stable-c} implies \eqref{eq:stochastic-d}.

\subsection*{Proof of Lemma~\ref{lemma:drift}}

Using constraint \eqref{eq:mean-a}, we have
\begin{align}
   & (\tilde{q}_{i (t + 1)}^F)^2 - (\tilde{q}_{i t}^F)^2 \nonumber \\
   =~ & \left( a_{i t}^F - \sum_{j \in \mathcal{J}} m_{i j t}^R \right)^2 + 2 \tilde{q}_{i t}^F \left( a_{i t}^F - \sum_{j \in \mathcal{J}} m_{i j t}^R \right) \nonumber \\
   \leq~ & \left( A_i^F + \sum_{j \in \mathcal{J}} M_{i j}^R \right)^2 + 2 \tilde{q}_{i t}^F \left( a_{i t}^F - \sum_{j \in \mathcal{J}} m_{i j t}^R \right). \nonumber
\end{align}
By constraint \eqref{eq:mean-f},
\begin{align}
   & (\tilde{q}_{t + 1}^E)^2 - (\tilde{q}_t^E)^2 \nonumber \\
   \leq~ & \left( \tilde{q}_t^E + \sum_{j \in \mathcal{J}} \gamma_{j t}^E p_{j t}^G - C^E \right)^2 - (\tilde{q}_t^E)^2 \nonumber \\
   =~ & \left( \sum_{j \in \mathcal{J}} \gamma_{j t}^E p_{j t}^G - C^E \right)^2 + 2 \tilde{q}_t^E \left( \sum_{j \in \mathcal{J}} \gamma_{j t}^E p_{j t}^G - C^E \right) \nonumber \\
   \leq~ & \left( \sum_{j \in \mathcal{J}} \overline{\gamma}_j^E \left( P_j^B + P_j^{SC} + P_j^{SD} + P_j^C \right) + C^E \right)^2 \nonumber \\
   & + 2 \tilde{q}_t^E \left( \sum_{j \in \mathcal{J}} \gamma_{j t}^E (p_{j t}^B + p_{j t}^{SC} - p_{j t}^{SD} + p_{j t}^C) - C^E \right). \nonumber
\end{align}
Other terms can be handled similarly. Then we have $\Delta_t \leq I + B$, where
\begin{align}
    B & = \sum_{i \in \mathcal{I}} \left( A_i^F + \sum_{j \in \mathcal{J}} M_{i j}^R \right)^2 + \sum_{j \in \mathcal{J}} \left( \sum_{i \in \mathcal{I}} M_{i j}^R + P_j^B \right)^2 \nonumber \\
    & + \sum_{j \in \mathcal{J}} \left(P_j^{SC} \eta_j^{SC} + P_j^{SD} / \eta_j^{SD}\right)^2 \nonumber \\
    & + \sum_{j \in \mathcal{J}} \left(\kappa_j^B P_j^B + \kappa_j^C P_j^C + \max \left\{ \left|\underline{\beta}_j^C\right|, \left|\overline{\beta}_j^C\right| \right\} \right)^2 \nonumber \\
    & + \left( \sum_{j \in \mathcal{J}} \overline{\gamma}_j^E \left( P_j^B + P_j^{SC} + P_j^{SD} + P_j^C \right) + C^E \right)^2. \nonumber
\end{align}

\subsection*{Proof of Theorem~\ref{thm:feasibility}}

To make the strategy by Algorithm~\ref{alg:online} feasible in problem \eqref{eq:stochastic}, we only need to ensure the following constraints:
\begin{subequations}
\begin{align}
    & \tilde{q}_{i t}^F - \theta_i^F \geq 0, \forall i \in \mathcal{I}, \tilde{q}_{j t}^B - \theta_j^B \geq 0, \forall j \in \mathcal{J}, \nonumber \\
    & \tilde{q}_{j t}^S - \theta_j^S \in \left[\underline{E}_j^S, \overline{E}_j^S\right],~ \tilde{q}_{j t}^H \in \left[ \underline{T}_j^H, \overline{T}_j^H \right], \forall j \in \mathcal{J}. \nonumber
\end{align}
\end{subequations}

According to the dynamic equations of these queues in \eqref{eq:mean-a}--\eqref{eq:mean-f}, the following constraints constitute a sufficient condition (note that the temperature cannot drop below the lower bound without the cooling system, according to Assumption~\ref{assumption:bound}; the cooling system is able to prevent temperature increase according to Assumption~\ref{assumption:QE}):
\begin{align}
    & \tilde{q}_{i t}^F - \theta_i^F \in \left[ 0, \sum_{j \in \mathcal{J}} M_{i j}^R \right) \implies m_{i j t}^{R *} = 0, \forall j, \nonumber \\
    & \tilde{q}_{j t}^B - \theta_j^B \in \left[ 0, P_j^B \right) \implies p_{j t}^{B *} = 0, \nonumber \\
    & \tilde{q}_{j t}^S - \theta_j^S \in \left[ \underline{E}_j^S, \underline{E}_j^S + P_j^{SD}/\eta_j^{SD} \right) \implies p_{j t}^{SD *} = 0, \nonumber \\
    & \tilde{q}_{j t}^S - \theta_j^S \in \left( \overline{E}_j^S - P_j^{SC} \eta_j^{SC}, \overline{E}_j^S \right] \implies p_{j t}^{SC *} = 0, \nonumber \\
    & \tilde{q}_{j t}^H - \theta_j^H \in \left[ \underline{T}_j^S, \underline{T}_j^S + \kappa_j^C P_j^C \right) \implies p_{j t}^{C *} = 0, \nonumber \\
    & \tilde{q}_{j t}^H - \theta_j^H \in \left( \overline{T}_j^S - \kappa_j^B P_j^B + \underline{\beta}_j^C, \overline{T}_j^S \right] \implies p_{j t}^{C *} = P_j^C. \nonumber 
\end{align}

The objective function of problem \eqref{eq:online} can be written as:
\begin{align}
    & \sum_{i \in \mathcal{I}} (\tilde{q}_{i t}^F - V \gamma_i^F) a_{i t}^F + \sum_{i \in \mathcal{I}} \sum_{j \in \mathcal{J}} (- \tilde{q}_{i t}^F + \tilde{q}_{j t}^B + V \gamma_{i j}^R) m_{i j t}^R \nonumber \\
    & + \sum_{j \in \mathcal{J}} (- \tilde{q}_{j t}^B + \tilde{q}_{j t}^H \kappa_j^B + \tilde{q}_t^E \gamma_{j t}^E + V \gamma_{j t}^G) p_{j t}^B \nonumber \\
    & + \sum_{j \in \mathcal{J}} (\tilde{q}_{j t}^S \eta_j^{SC} + \tilde{q}_t^E \gamma_{j t}^E + V \gamma_j^S + V \gamma_{j t}^G) p_{j t}^{SC} \nonumber \\
    & + \sum_{j \in \mathcal{J}} (- \tilde{q}_{j t}^S / \eta_j^{SD} - \tilde{q}_t^E \gamma_{j t}^E + V \gamma_j^S - V \gamma_{j t}^G) p_{j t}^{SD} \nonumber \\
    & + \sum_{j \in \mathcal{J}} (- \tilde{q}_{j t}^H \kappa_j^C + \tilde{q}_t^E \gamma_{j t}^E + V \gamma_{j t}^G) p_{j t}^C \nonumber \\
    & - \tilde{q}_t^E C^E + V \sum_{i \in \mathcal{I}} \gamma_i^F \alpha_{i t}^F. \nonumber 
\end{align}
Therefore, we only need the following conditions to ensure feasibility:
\begin{subequations}
\begin{align}
    & \tilde{q}_{i t}^F - \theta_i^F < \sum_j M_{i j}^R \implies - \tilde{q}_{i t}^F + \tilde{q}_{j t}^B + V \gamma_{i j}^R > 0, \forall j, \nonumber \\
    & \tilde{q}_{j t}^B - \theta_j^B < P_j^B \implies - \tilde{q}_{j t}^B + \tilde{q}_{j t}^H \kappa_j^B + \tilde{q}_t^E \gamma_{j t}^E + V \gamma_{j t}^G > 0, \nonumber \\
    & \tilde{q}_{j t}^S - \theta_j^S < \underline{E}_j^S + P_j^{SD}/\eta_j^{SD} \implies \nonumber \\
    & - \tilde{q}_{j t}^S / \eta_j^{SD} - \tilde{q}_t^E \gamma_{j t}^E + V \gamma_j^S - V \gamma_{j t}^G > 0, \nonumber \\
    & \tilde{q}_{j t}^S - \theta_j^S > \overline{E}_j^S - P_j^{SC} \eta_j^{SC} \implies \nonumber \\
    & \tilde{q}_{j t}^S \eta_j^{SC} + \tilde{q}_t^E \gamma_{j t}^E + V \gamma_j^S + V \gamma_{j t}^G > 0, \nonumber \\
    & \tilde{q}_{j t}^H - \theta_j^H < \underline{T}_j^S + \kappa_j^C P_j^C \implies \nonumber \\
    & - \tilde{q}_{j t}^H \kappa_j^C + \tilde{q}_t^E \gamma_{j t}^E + V \gamma_{j t}^G > 0, \nonumber \\
    & \tilde{q}_{j t}^H - \theta_j^H < \overline{T}_j^S - \kappa_j^B P_j^B + \underline{\beta}_j^C \implies \nonumber \nonumber \\
    & - \tilde{q}_{j t}^H \kappa_j^C + \tilde{q}_t^E \gamma_{j t}^E + V \gamma_{j t}^G > 0, \nonumber 
\end{align}
\end{subequations}
which can be guaranteed by \eqref{eq:requirement} according to the bounds of uncertainties and queues.

\end{multicols}

\end{document}